\documentclass[cits]{PoS}

\usepackage{bbm}
\usepackage{amsmath}
\usepackage{amssymb}

\newcommand{\ee}{{\mathrm{e}}}

\newcommand{\eucl}{{\mathrm{E}}}
\newcommand{\al}{\alpha}
\newcommand{\be}{\beta}

\title{A Variational Study of the Nucleon Wavefunction }

\ShortTitle{A Variational Study of the Nucleon Wavefunction }

\author{\speaker{Jonathan D. Bratt}\thanks{E-mail: jdbratt@mit.edu}  \, and J.W.~Negele  \thanks{E-mail: negele@mit.edu}\\
        Center for Theoretical Physics, Massachusetts Institute of Technology, Cambridge, MA 02139 \\
 }

\abstract{The structure of the nucleon is studied variationally on the lattice by maximizing the overlap between the nucleon and 
a trial function generated by an interpolating field containing variational parameters.  We examine the effect of the spatial extent of  smeared quark sources, the degree of gauge field smearing,  the positions of smeared quark sources, inclusion of lower Dirac components and of diquark substructure.   Exploratory calculations  with 
quenched Wilson fermions at a pion mass of 900 MeV achieved overlaps as high as 80\%, and there was no evidence of diquark substructure.}

\FullConference{The XXVI International Symposium on Lattice Field Theory \\
                 July 14 - 19, 2008\\
                 Williamsburg, Virginia, USA}

\begin{document}

\section{Introduction}

Experimental studies of nucleon structure have revealed many features of form factors, parton distributions, and  generalized form factors to impressive precision, and lattice calculations are now beginning to calculate these  observables from first principles in agreement with experiment.  However, a complicated numerical calculation that merely reproduces an experimental number does not necessarily, in itself, provide any physical insight as to why and how QCD works to produce a given electric or magnetic form factor or a particular decomposition of the nucleon spin. Hence, it is desirable to use lattice QCD as a tool to obtain insight into the quark and gluon structure of the nucleon.

Calculation of a  nucleon on the lattice occurs in two steps.  First, we create a state, which we will refer to as a trial state, by acting with an interpolating field with the proper quantum numbers on the QCD vacuum.  Then, that state is evolved in Euclidean time to project out the physical ground state. Since, as described below, it is straightforward to extract the overlap between normalized nucleon and trial states, we can use the overlap as a quantitative criterion to systematically improve the trial state and thereby obtain insight into key features of the nucleon wave function.

Since the only practical way to calculate nucleon correlation functions on a lattice is to express them in terms of quark correlation function, the choices in sources are the positions of each quark, the spatial extent of the  quark source achieved by gauge invariant smearing, the degree of smoothing of gluon fluctuations achieved by smearing the gauge links included in the source, the inclusion of only upper or of both upper and lower Dirac components, and the choice of $\gamma$ matrix structure used to produce the nucleon quantum numbers.

Given recent resurgence of interest\cite{Jaffe:2003sg,jaffe,Wilczek:2004im} in diquarks \cite{ida}, it is of interest to look for any evidence of diquark correlations variationally.  We could, in principle, study  two possible diquark configurations in a nucleon\cite{ioffe, alexandrou}, the scalar channel $(u\, C \gamma_5 \,d)$, and the vector channel $(u\, C \gamma_\mu \, d) $. However, at the one gluon exchange level, quarks in the scalar configuration have lower energy than quarks in the vector configuration \cite{jaffe}, which are thus called "good" and "bad" diquarks respectively, and we will focus our attention only on the good diquarks and consider sources of the form $(u\, C \gamma_5 \, d)\, u$.  Analogous to the simple picture of a meson comprised of an antiquark and quark connected by a flux tube, which leads naturally to Regge trajectories, in the limit in which two quarks bound into a point-like diquark in a nucleon, one could also develop a ``dog bone'' model of baryons with a diquark and quark connected by a flux tube. This picture is phenomenologically successful(Wilczek).  We can explore this physics in our trial function, by allowing for a diquark to have a different degree of spatial localization and to be separated from the remaining quark.

\section{Variational Method}
In this study, we used a simple variational approach to study the ground-state wavefunction of the nucleon. For a trial source operator $J(x)$, the overlap with the ground state is calculated from a fit to the nucleon two-point correlation function \cite{Dolgov:2002zm}. 
Starting with the correlation function in position space:

\begin{displaymath}
C(\mathbf{x},t) = \langle \Omega | \bar{J}(\mathbf{x},t) J(\mathbf{0},0) | \Omega \rangle 
\end{displaymath}
we insert a complete set of states and project onto zero momentum in the usual way to obtain:
$$
C(t) \equiv  \sum_\mathbf{x} C(\mathbf{x},t) 
=  \sum_n e^{-E_n t} | \langle n | J(\mathbf{0},0) | \Omega \rangle |^2
\equiv  \sum_n A_n e^{-E_n t} .
$$

The energies $E_n$ and coefficients $A_n$ are extracted by fitting $C(t)$ to a sum of exponentials, and the normalized overlap of our trial source with the nucleon ground state is given by:

\begin{equation}
\frac{| \langle 0 | J(0) | \Omega \rangle |^2 }{ \sum_n | \langle n | J(0) | \Omega \rangle |^2 } = \frac{A_0}{ C(0) }
\label{overlap}
\end{equation}

Note that this variational study with a single source differs from the variational approach used extensively in spectroscopy\cite{Michael:1985ne}, which considers a superposition of  distinct sources with arbitrary coefficients and determines the optimal coefficients by  minimizing the energy, rather than maximizing the overlap.

\subsection{Correction at $t=0$}

For $t>0$, the  correlator  $C(t) = \langle \Omega |\hat{\bar J}(t) \hat{J}(0) |\Omega \rangle$
is equal to the transfer matrix element
\begin{equation}
  \label{eq:Ccan}
  C(t) = \langle\Omega | \hat{\bar J} \ee^{-\hat H t} \hat{J} |\Omega\rangle
  \;.
\end{equation}

For $t=0$, the Euclidean correlator is no longer equal to the simple
matrix element (\ref{eq:Ccan}), but rather to the vacuum expectation
value of the normal-ordered product of $ \hat{\bar J}$ and $ \hat{J}$~\cite{Luscher},
\begin{equation}
\label{cf-this}
  C^\eucl_{i j}(0)
  = \langle\Omega|N[\hat{\bar J}  \hat{J}]|\Omega\rangle
\;,
\end{equation}
where normal ordering is defined by its action on quark
operators: in a basis where \newline
$\gamma_0= \rm{diag}(1,1,-1,-1)$,
\begin{equation}\label{eq:NO}
  N[\hat q_{a \alpha}(\vec x) \hat{\bar q}_{b \beta}(\vec y)]
  = \begin{cases}
    \hat q_{a \gamma}(\vec x) \hat{\bar q}_{b \beta}(\vec y)
    & \text{if $\alpha,\beta=1,2$} \\
    - \hat{\bar q}_{b \beta}(\vec y) \hat q_{a \gamma}(\vec x)
    & \text{if $\alpha,\beta=3,4$}
\;.
  \end{cases}
\end{equation}
In order to compute the matrix element (\ref{eq:Ccan}) without
normal-ordering, we have to sum the Euclidean expectation value over
all contractions.  In practice, this is achieved by replacing the
na\"{i}ve quark propagator at equal time by the one corresponding to a
non-normal-ordered expectation value\cite{Jahn,Sigaev},
\begin{eqnarray}
  S'_{a i\al,b j\be}(\vec x,t;\vec y,t')
  &=& S_{a i\al,b j\be}(\vec x,t;\vec y,t')\nonumber \\
  &+& C_{a i\al,b j\be}(\vec x,t;\vec y,t')\;,
\end{eqnarray}
where
\begin{eqnarray}
  C_{a i\al,b j\be}(\vec x,t;\vec y,t') &=&
  \hat q_{a i\al}(\vec x,t) \hat{\bar q}_{b j\be}(\vec y,t')\nonumber\\
  &-& N[\hat q_{a i\al}(\vec x,t) \hat{\bar q}_{b j\be}(\vec y,t')]
\end{eqnarray}
depends on the fermion action.  For the Wilson action~\cite{Luscher},
\begin{equation}
  C_{a i\al,b j\be}(\vec x,t;\vec y,t') =
  - \delta_{i j} (1-\gamma_4)_{\al\be} B^{-1}_{a b}(\vec x,\vec y)
  \delta(t,t')
\end{equation}
with
\begin{eqnarray}
\label{BBB}
  B_{a b}(\vec x,\vec y)= \delta_{a b} \delta(\vec x,\vec y) \hspace{4.2cm}\nonumber \\
 \quad - K \sum_{i=1}^3 \left[ 
    U_{i}^{a b}(\vec x) \delta(\vec x+\hat\imath,\vec y)
    + U^{b a *}_{i}(\vec y) \delta(\vec x-\hat\imath,\vec y)
  \right].
\end{eqnarray}

In our calculations with the Wilson action, we include this term to calculate the correct value of $C(0)$. However, there is no comparable transfer matrix for domain wall fermions, so we cannot presently extend this work to chiral fermions.

\section{This Calculation}

\begin{figure}
\begin{minipage}[t]{0.5\linewidth}
\begin{center}
\leavevmode
\includegraphics[width=1\textwidth]{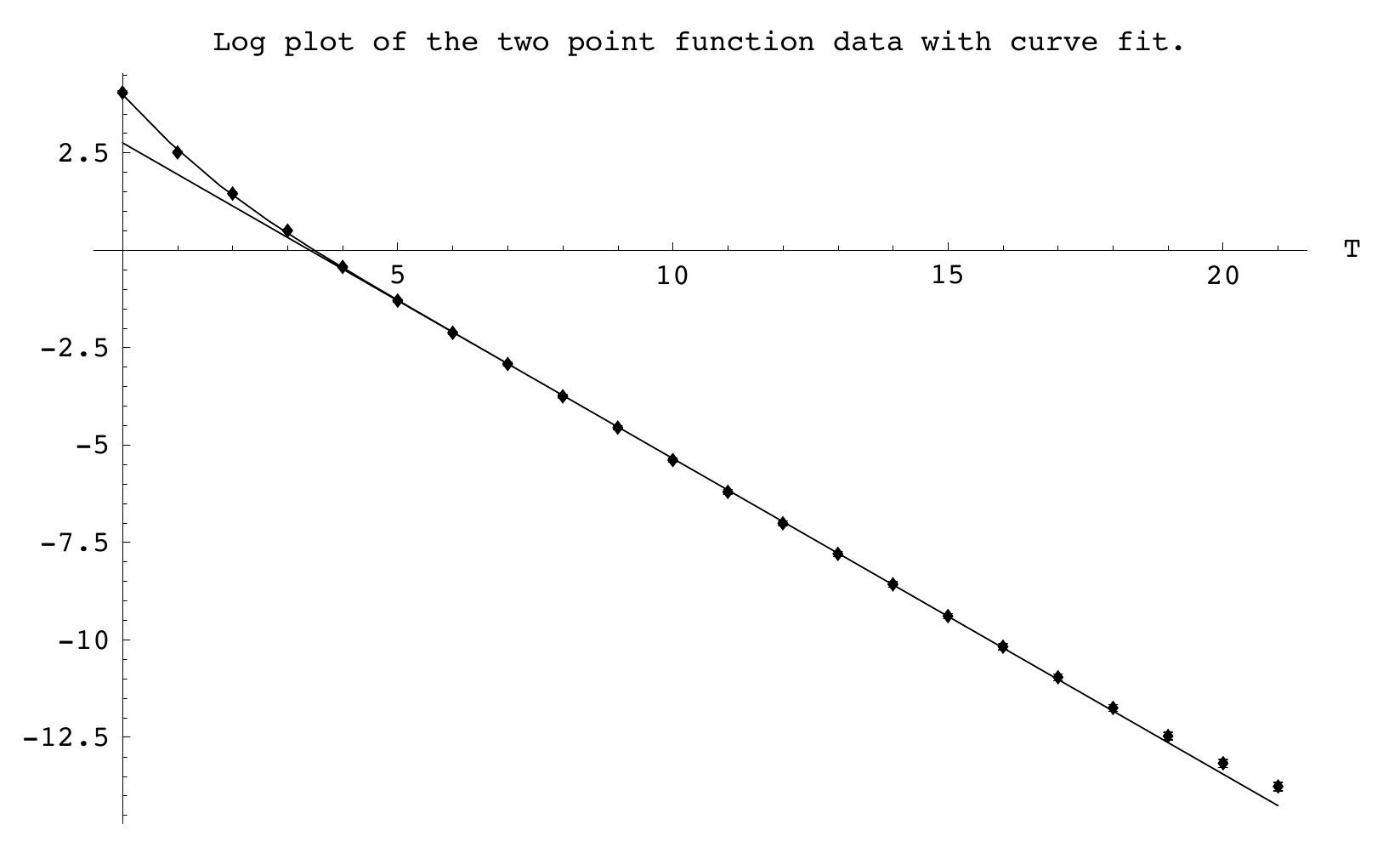}
\end{center}
\caption{Plot of a typical two-point correlation function.}
\label{fig:twopoint}
\end{minipage}
\hspace{0.5cm} 
\begin{minipage}[t]{0.5\linewidth} 
\begin{center}
\leavevmode
\includegraphics[width=1\textwidth]{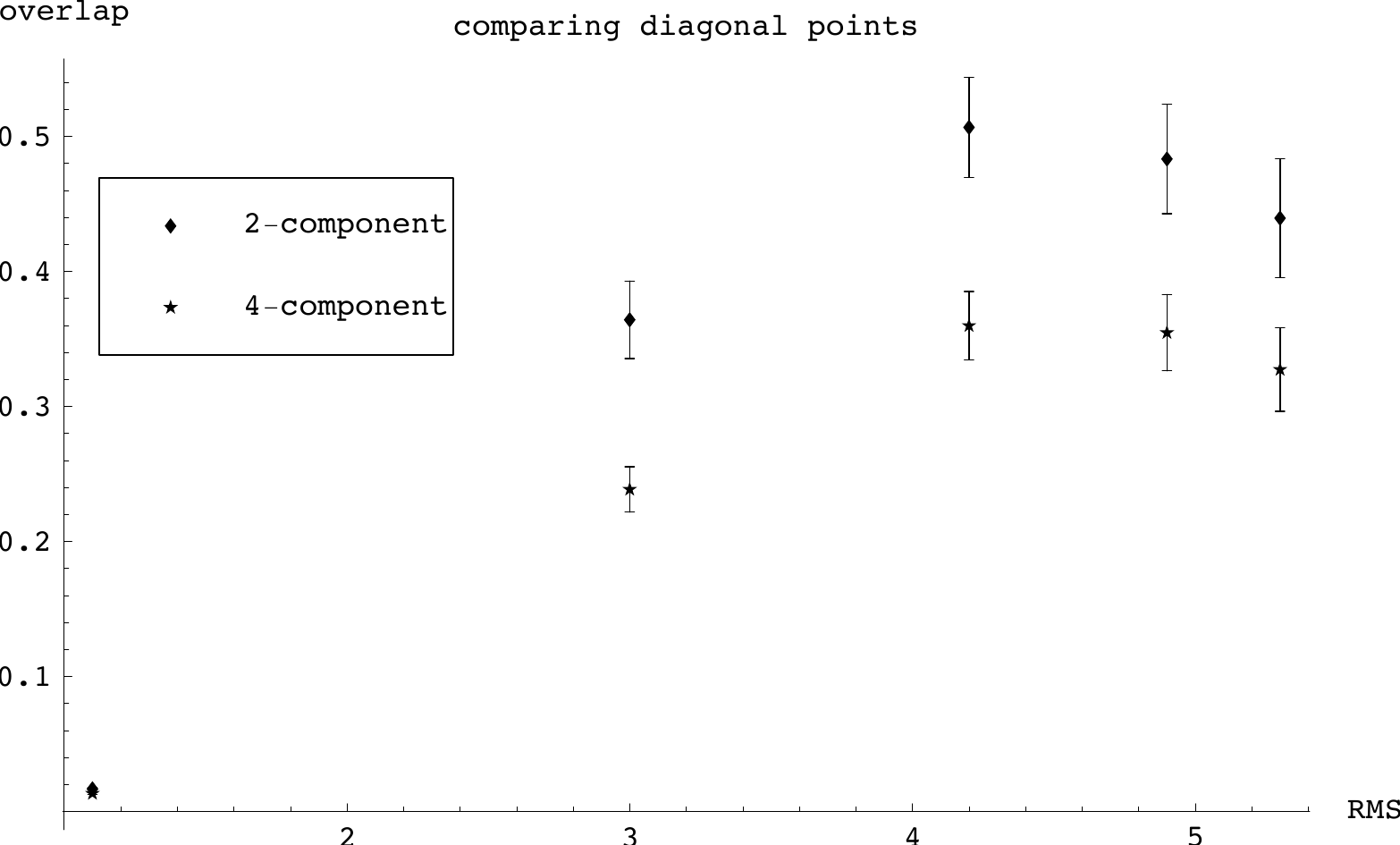}
\end{center}
\caption{Overlap as a function of quark RMS radius (in lattice units). Results are shown for both two- and four-component spinors.  }
\label{fig:twovfour}
\end{minipage}
\end{figure}

We performed our exploratory calculation  on $16^3 \times 32 $  quenched Wilson lattices with $\beta = 6.0$ and $\kappa = 0.1530$ ($m_{\pi} \approx 900 MeV$) for ensemble sizes of 100 - 200 configurations.  The trial sources were of the form 
\begin{equation}
J_\alpha  = \big( U_\beta \; [C \, \gamma_5]_{\beta \gamma} \; D_\gamma \big) \;  U_\alpha.
\end{equation}

We varied the number of gauge invariant smearing steps for the quark sources, controlling their RMS radius\cite{Dolgov:2002zm} , the number of dirac spinor components (four, or two in the non-relativistic limit), the gauge field smearing in the source links, the relative size of quark and  diquark smearing, and the relative position of the quark and diquark.

\subsection{Results}

\begin{figure}
\begin{minipage}[t]{0.5\linewidth}
\begin{center}
\leavevmode
\includegraphics[width=1\textwidth]{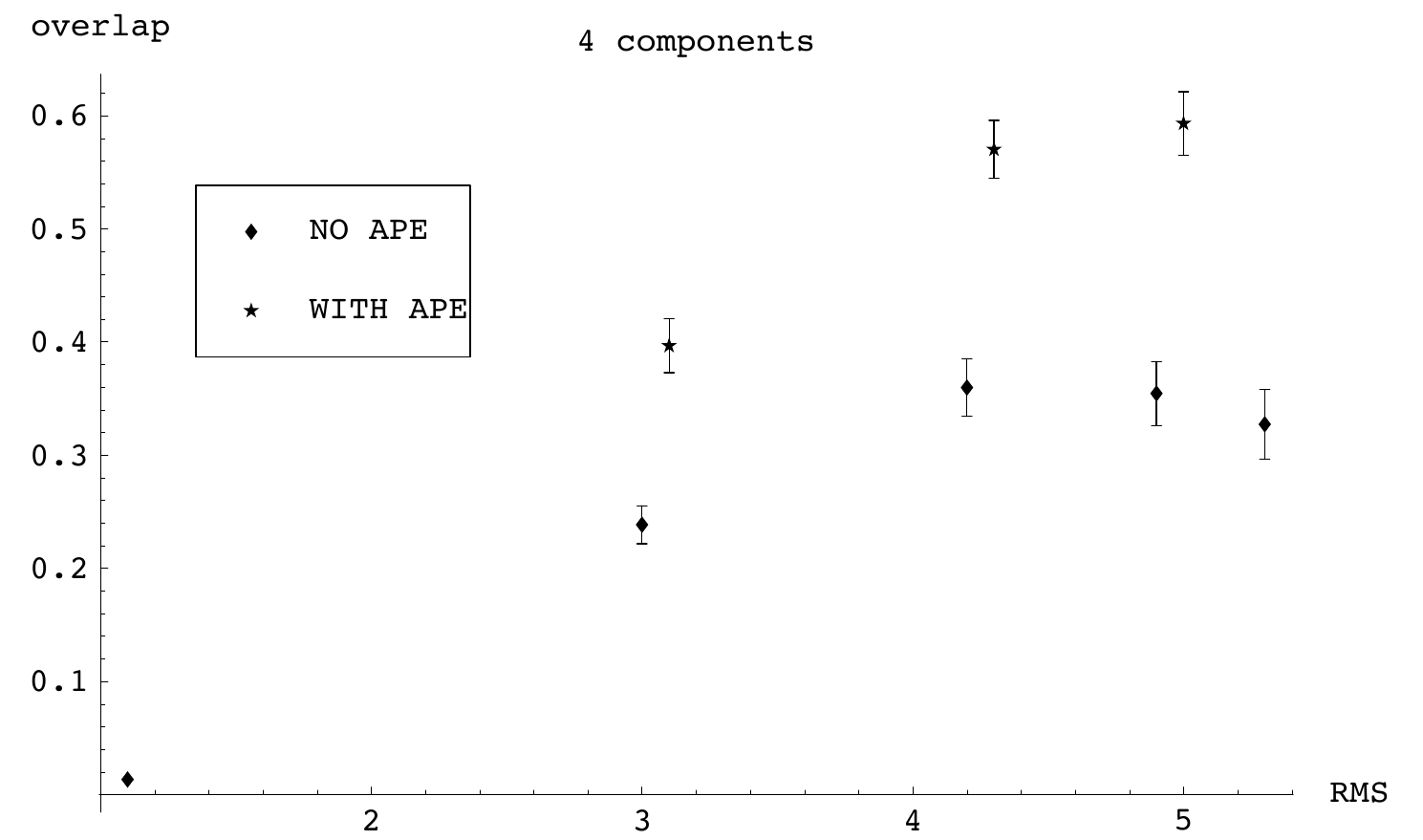}
\end{center}
\caption{Overlap as a function of quark RMS radius for four-component spinors. Results are shown for calculations with and without APE smearing.}
\label{fig:ape4}
\end{minipage}
\hspace{0.5cm} 
\begin{minipage}[t]{0.5\linewidth}
\begin{center}
\leavevmode
\includegraphics[width=1\textwidth]{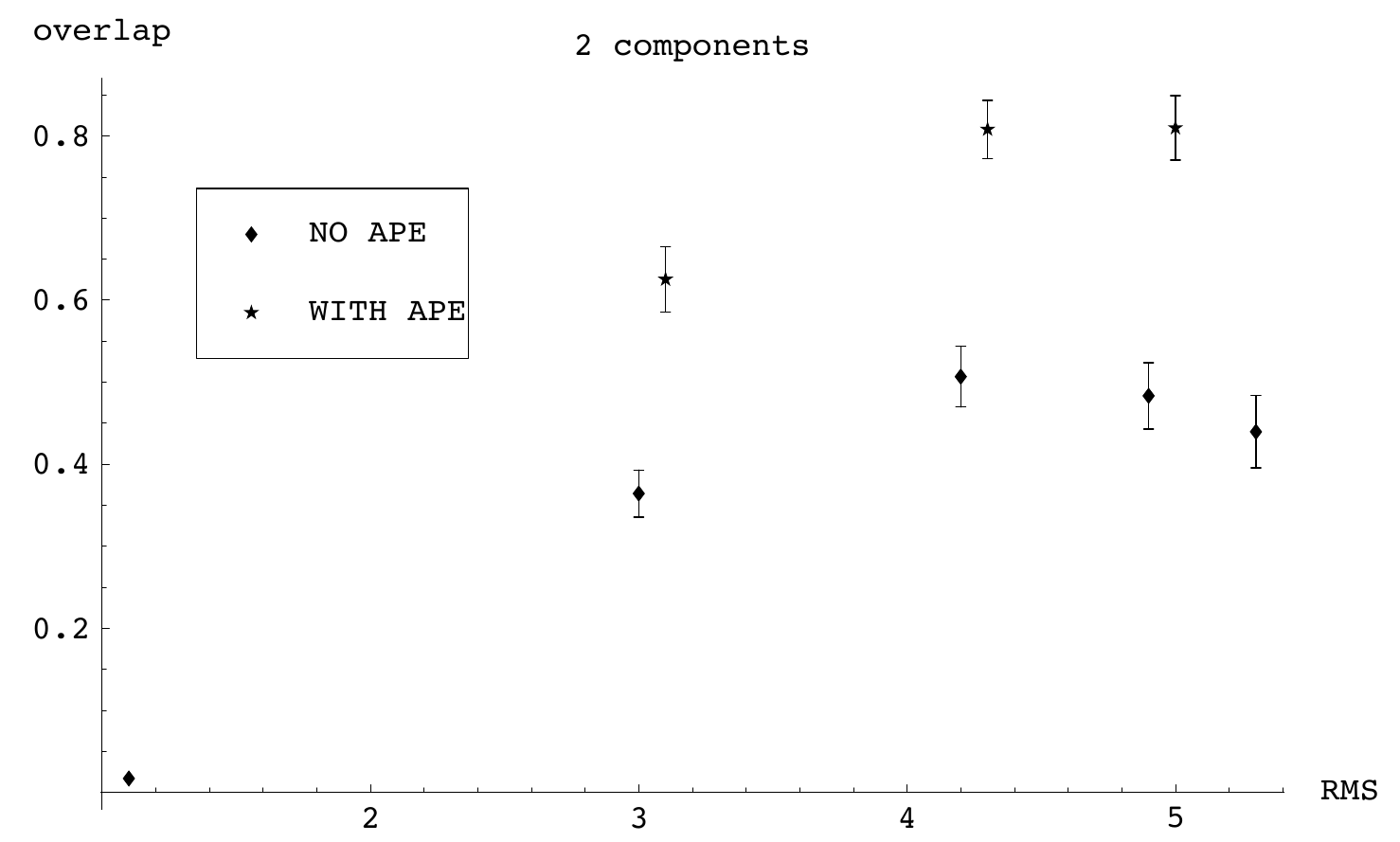}
\end{center}
\caption{Overlap as a function of quark RMS radius for two-component spinors. Results are shown for calculations with and without APE smearing.}
\label{fig:ape2}
\end{minipage}
\end{figure}

\begin{figure}
\begin{minipage}[t]{0.5\linewidth} 
\begin{center}
\leavevmode
\includegraphics[width=1\textwidth]{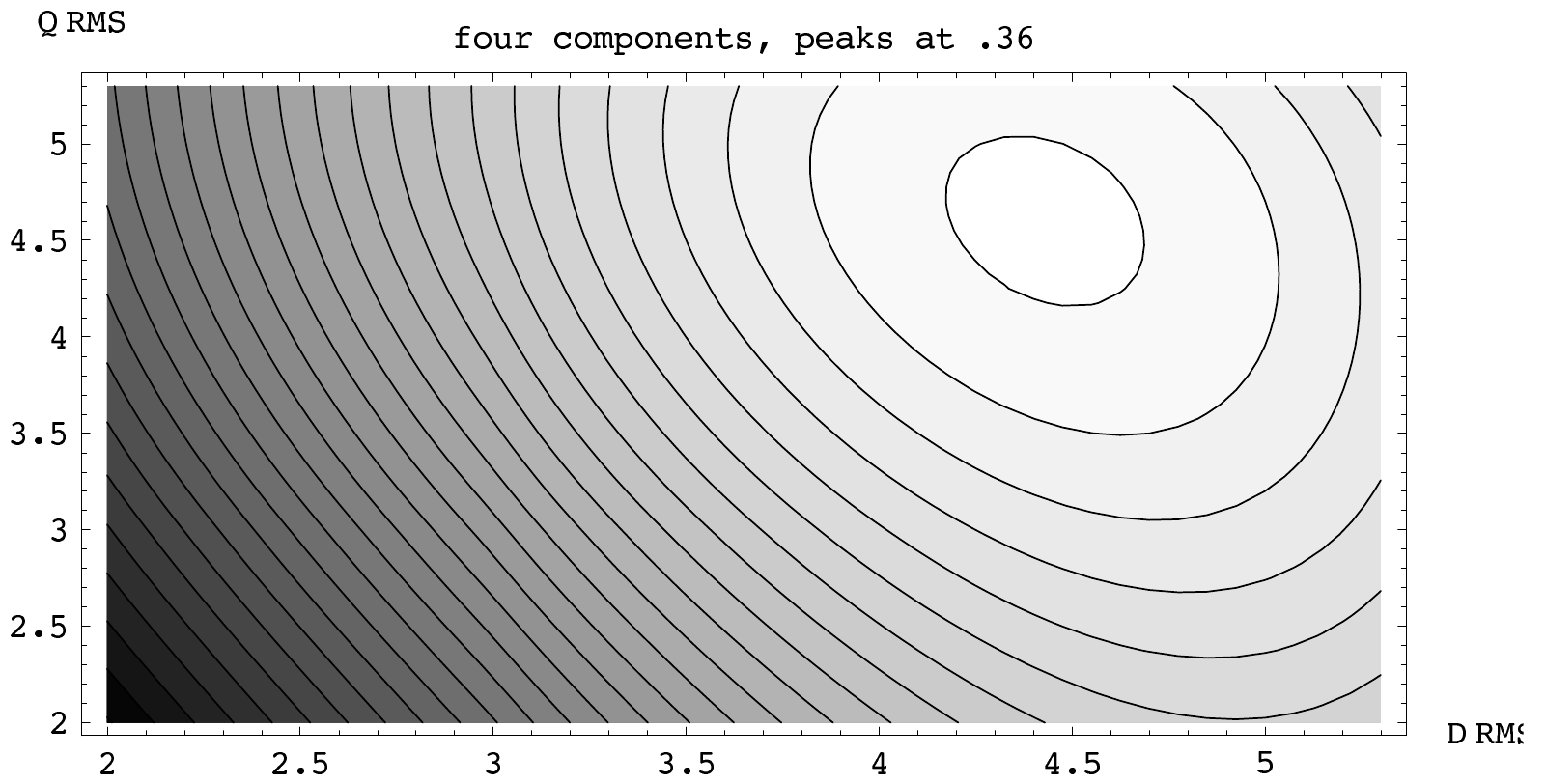}
\end{center}
\caption{Contour plot of the overlap as a function of both quark and diquark RMS radius, for four-component spinors.}
\label{fig:contour4}
\end{minipage}
\hspace{0.5cm}
\begin{minipage}[t]{0.5\linewidth}
\begin{center}
\leavevmode
\includegraphics[width=1\textwidth]{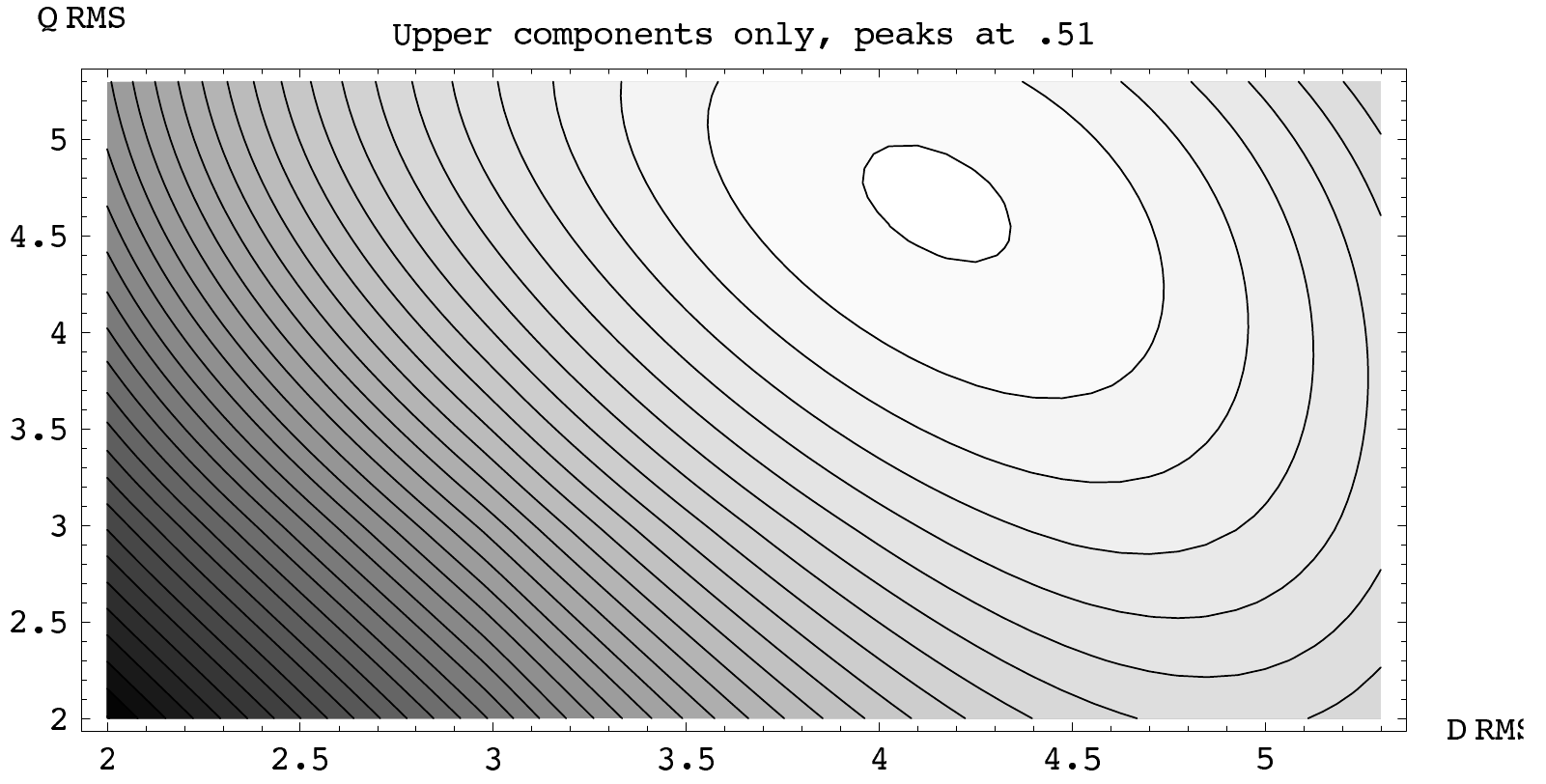}
\end{center}
\caption{Contour plot of the overlap as a function of both quark and diquark RMS radius, for two-component spinors.}
\label{fig:contour2}
\end{minipage}
\end{figure}

\begin{figure}
\begin{minipage}[t]{0.5\linewidth} 
\begin{center}
\leavevmode
\includegraphics[width=1\textwidth]{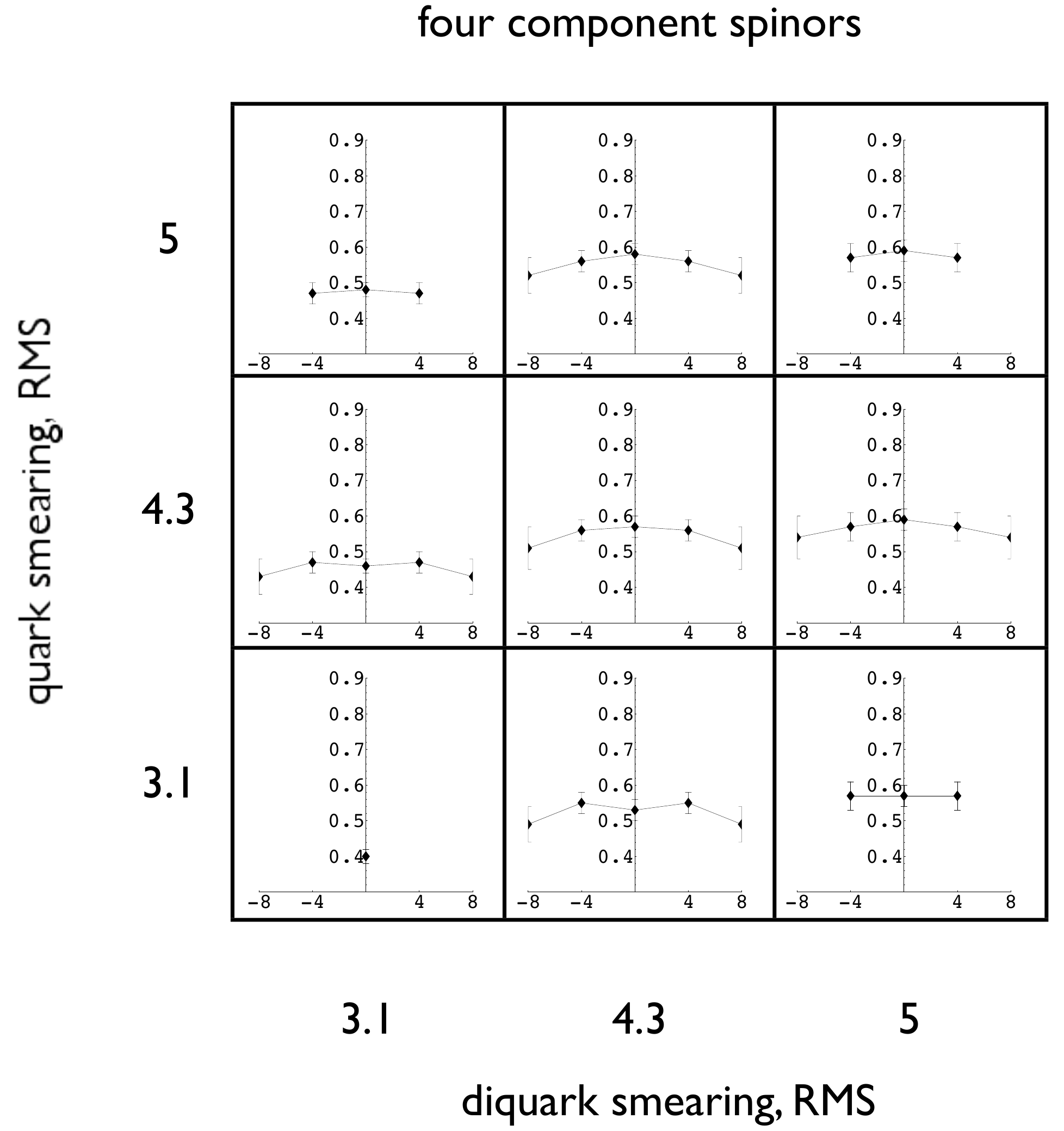}
\end{center}
\caption{Array of plots of overlap as a function of quark displacement, for four-component spinors.}
\label{fig:array4}
\end{minipage}
\hspace{0.9cm} 
\begin{minipage}[t]{0.5\linewidth}
\begin{center}
\leavevmode
\includegraphics[width=1\textwidth]{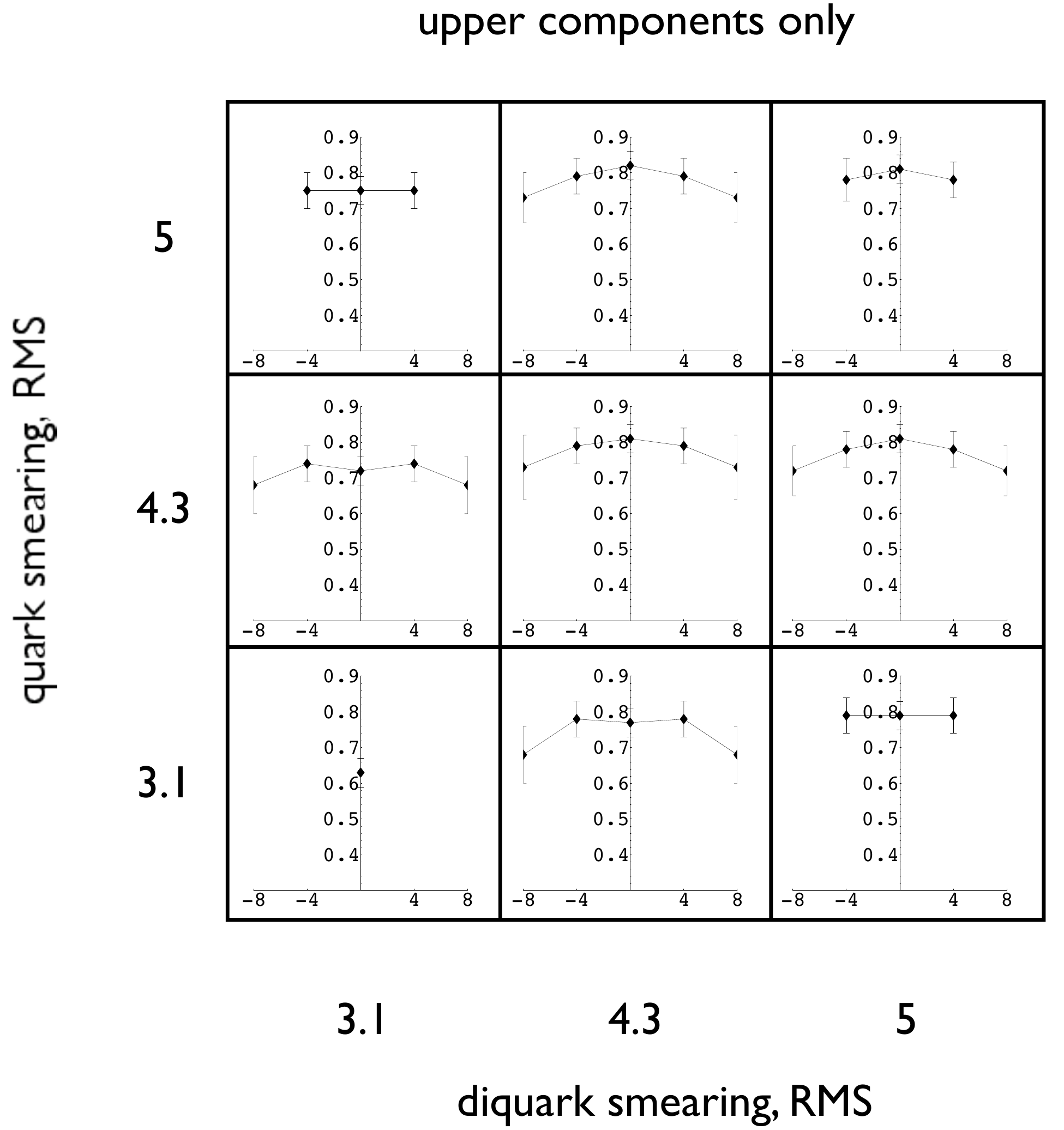}
\end{center}
\caption{Array of plots of overlap as a function of quark displacement, for two-component spinors.}
\label{fig:array2}
\end{minipage}
\end{figure}

\smallskip
\noindent{ \bf Quark Smearing}

We smear all three quark fields to the same RMS radius and locate them at the same spatial position.
Results for the variation of the quark RMS radius are shown in Fig. \ref{fig:twovfour}. The overlap behaves as in previous studies, starting at the order of $10^{-5}$  for a small quark radius, increasing to a maximum at some finite radius, and falling off at larger radii. This reflects the finite spatial extent of the quark wavefunction within a nucleon. The peak occurs around $4.5$ lattice units ($\approx 0.4$ fm), consistent with \cite{Dolgov:2002zm}.

\smallskip
\noindent{ \bf Two vs Four Spinor Components}

When constructing the nucleon two-point function, we can use either the full quark spinor $q$, or the projected spinor $\frac{1+\gamma_0}{2}q$, which corresponds to the upper two components in the Dirac basis. We observe (Fig. \ref{fig:twovfour}) that using the two-component spinors increases the overlap significantly. This is consistent with the expectation that the lower components be in a p-wave type state, which has very poor overlap with the approximately gaussian wavefunction used as a trial source (also see discussion in \cite{gockeler}).

\smallskip
\noindent{ \bf Gauge Field Smearing}

Because of the large gluon fluctuations associated with the link variables, the overlap is increased by 
smearing the gauge links used in constructing the quark sources. For this, we used 25 iterations of APE smearing  with $\epsilon = 0.35$ in the notation of \cite{albanese}. Figs. \ref{fig:ape4} and \ref{fig:ape2} show the overlap calculated with APE smeared links compared with the overlap for no link smearing. We note that smearing the gauge fields significantly improves the calculated overlaps.

\smallskip
\noindent{ \bf Relative Size of Quark, Diquark}

Using different fermion smearing parameters for the diquark and lone quark parts of the nucleon source operator allows us to explore trial sources with different spatial sizes for the diquark and the remaining quark. Contour plots of the resulting two-dimensional parameter space are shown in Figs.  \ref{fig:contour4} and  \ref{fig:contour2}. The primary feature of these results is the location of the peak overlap (very nearly) along the previously explored diagonal cross-section. Since any significant asymmetry in the peak would have been suggestive of a strong diquark substructure, this is one piece of evidence against the presence of large diquark correlations.

\smallskip
\noindent{ \bf Relative Displacement of Quark, Diquark}

To explore the physical ``dog bone picture'' of a separated quark and diquark joined by a flux tube, we also tried displacing the single quark from the diquark in our trial sources. Results for a variety of displacement lengths and smearing combinations are shown in Figs.  \ref{fig:array4} and  \ref{fig:array2}. The maximum overlap is observed for zero displacement, again suggesting the lack of strong diquark correlations.

\section{Conclusions}

In summary, we observe dramatic increases in the overlap between a trial state and the physical nucleon as we vary accessible features of the trial state.  Smearing the quark interpolating fields from a point to an optimal RMS radius increases the overlap for a four-component trial function from a fraction of a percent to 35\%.  Removing the unphysical S-wave lower components increases the overlap from 35\% to 50\%.  Optimal   smearing of the gauge field further improves the overlap from 50\% to more than 80\%
Attempts to further increase the overlap by including diquark correlations associated with the relative size and position of the quarks and diquarks yielded no further improvement, indicating that such correlations are not significantly favored in the nucleon ground state.  

From the perspective of lattice calculation technology, these results are extremely useful in generating sources that involve minimal contaminants from excited states.  From a physics perspective, they give useful insight into both what the quark and gluon degrees of freedom are doing and what they are not doing.  It would be valuable to extend these calculations to lighter quark masses, but for our research exploiting light domain wall quarks, we need to find an alternative to the transfer matrix corrections described in this work.

\section*{Acknowledgements}

This work was supported in part  by the DOE Office of Nuclear Physics under grant
DE-FG02-94ER40818. 
and used computer resources provided by the DOE through  its support of the MIT Blue Gene/L under 
grant DE-FG02-05ER25681.


\begin{thebibliography}{99}


  
%\cite{Jaffe:2003sg}
\bibitem{Jaffe:2003sg}
  R.~L.~Jaffe and F.~Wilczek,
  %``Diquarks and exotic spectroscopy,''
  Phys.\ Rev.\ Lett.\  {\bf 91}, 232003 (2003)
  [arXiv:hep-ph/0307341].
  %%CITATION = PRLTA,91,232003;%%


\bibitem{jaffe}
R. L. Jaffe, 
Phys. Rept. 409, 1 (2005).

  
%\cite{Wilczek:2004im}
\bibitem{Wilczek:2004im}
  F.~Wilczek,
  %``Diquarks as inspiration and as objects,''
  arXiv:hep-ph/0409168.
  %%CITATION = HEP-PH/0409168;%%
  

\bibitem{ida} 
M. Ida and R. Kobayashi,
 Prog. Theor. Phys. 36, 846 (1966).


\bibitem{ioffe}
B.L.~Ioffe,
Nucl. Phys. B188, 317 (1981).


\bibitem{alexandrou}
C. Alexandrou, Ph. de Focrand and B. Lucini,
Phys. Rev. Lett. \textbf{97}, 222002 (2006).


%\cite{Dolgov:2002zm}
\bibitem{Dolgov:2002zm}
  D.~Dolgov {\it et al.}  [LHPC collaboration and TXL Collaboration],
  %``Moments of nucleon light cone quark distributions calculated in full
  %lattice QCD,''
  Phys.\ Rev.\  D {\bf 66}, 034506 (2002)
  [arXiv:hep-lat/0201021].
  %%CITATION = PHRVA,D66,034506;%%
  
 
  
%\cite{Michael:1985ne}
\bibitem{Michael:1985ne}
  C.~Michael,
  %``Adjoint Sources In Lattice Gauge Theory,''
  Nucl.\ Phys.\  B {\bf 259}, 58 (1985).
  %%CITATION = NUPHA,B259,58;%%
  


\bibitem{Luscher}
M. L\"uscher,
Commun. Math. Phys. 54, 283-292 (1977).

\bibitem{Jahn}
O. Jahn, private communication.

\bibitem{Sigaev}
D. Sigaev, MIT Ph.D. Thesis, (2008)

\bibitem{gockeler}
M. G\"ockeler \emph{et al.}, 
Nucl. Phys. Proc. Suppl. 42:337-345 (1995). 

\bibitem{albanese}
M. Albanese \emph{et al.}  [APE Collaboration], 
Phys. Lett. B 192, 163 (1987).

\end{thebibliography}
\end{document}